\documentclass[a4paper]{jpconf}
\usepackage{graphicx}

\begin{document}
\title{The Sun-as-a-star observations: GOLF \& VIRGO on SoHO, and BiSON network}

\author{R. A. Garc\'\i a$^1$, G.R. Davies$^{1,2}$, A. Jim\'enez$^{3,4}$, J. Ballot$^{5,6}$, S.~Mathur$^{7,8}$, D. Salabert$^{9}$, W. J. Chaplin$^2$, Y. Elsworth$^2$, C. R\'egulo$^{3,4}$, S. Turck-Chi\`eze$^1$}

\address{$^1$ Laboratoire AIM, CEA/DSM-CNRS-Universit\'e Paris Diderot; CEA, IRFU, SAp, F-91191, Gif-sur-Yvette, France}
\address{$^2$ School of Physics and Astronomy, University of Birmingham, Edgbaston, Birmingham, B15 2TT, UK}
\address{$^3$ Instituto de Astrof\'isica de Canarias, E-38200 La Laguna, Tenerife, Spain}
\address{$^4$ Dept. de Astrof\'isica, Universidad de La Laguna, E-38206 La Laguna, Tenerife, Spain}
\address{$^5$  CNRS, Institut de Recherche en Astrophysique et Plan\'etologie, 14 avenue Edouard Belin, 31400 Toulouse, France}
\address{$^6$ Universit\'e de Toulouse, UPS-OMP, IRAP, 31400 Toulouse, France}
\address{$^7$ High Altitude Observatory, 3080 Center Green Drive, Boulder, CO, 80302 USA}
\address{$^8$ Space Science Institute, 4750 Walnut Street, Suite 205, Boulder, Colorado 80301 USA}
\address{$^9$ Laboratoire Lagrange, UMR7293, Universit\'e de Nice Sophia-Antipolis, CNRS, Observatoire de la C\^ote d'Azur, Bd. de l'Observatoire, 06304 Nice, France}

\ead{rgarcia@cea.fr, davies@bison.ph.bham.ac.uk, ajm@iac.es, Jerome.ballot@irap.omp.eu, smathur@SpaceScience.org, salabert@oca.eu, wjc@bison.ph.bham.ac.uk, ype@bison.ph.bham.ac.uk, crr@iac.es, sylvaine.turck-chieze@cea.fr}

\begin{abstract}

Sun-as-a-star observations are very important for the study  of the conditions within the Sun and in particular for the deep interior where higher degree modes do not penetrate. They are also of significance in this era of dramatic advances in stellar asteroseismology as they are comparable to those measured in other stars by asteroseismic missions such as CoRoT, {\it Kepler}, and MOST. More than 17 years of continuous measurements of SoHO and more than 30 years of BiSON observations provide very long data sets of uninterrupted helioseismic observations. In this work, we discuss the present status of all these facilities that continue to provide state-of-the-art measurements and invaluable data to improve our knowledge of the deepest layers of the Sun and its structural changes during the activity cycle.Ê
\end{abstract}

\section{Introduction}
Helioseismology entered in a new era in the mid nineties with the deployment of the ground-based networks around the world such as the Birmingham Solar Oscillation Network (BiSON, \cite{Cha96}), the Global Oscillation Network Group (GONG, \cite{Har96}), and the launch of the Solar and Heliospheric Observatory (SoHO) mission \cite{Dom95}. All these facilities allowed a near continuous monitoring of the Sun, drastically improving the quality of the datasets available. 

By measuring the Sun as a star, i.e., without spatial resolution, we can access low-degree modes. These modes are particularly interesting because they reach the deepest regions of the Sun providing us with invaluable information of its structure (e.g. \cite{STC01,Mat07, Bas09}), and dynamics (e.g.\cite{Els95,STC09}). They have also been used very recently as new indicators of activity of the sub-surface layers \cite{Sal09,Sim13}. Both, SoHO and BiSON, will be in operation until 2016 ensuring the continuity of the activity-cycle studies as well as progressing in the detection of low-degree low-order p modes and hopefully g modes.

\section{SoHO Instruments: GOLF and VIRGO/SPM}
The ESA/NASA SoHO mission was launched on December 2nd 1995. SoHO carries on two of such instruments: The Global Oscillations at Low Frequency (GOLF, \cite{Gab95}) and the Variability of solar IRradiance and Gravity Oscillations (VIRGO, \cite{Fro95}). The overall duty cycle of these instruments is above 95\% for the last ~17 years. Most of the missing data correspond to the period July - October 2008 due to the loss and recovery of the spacecraft. 

\begin{figure}[!htb]
\includegraphics[width = 0.6\textwidth]{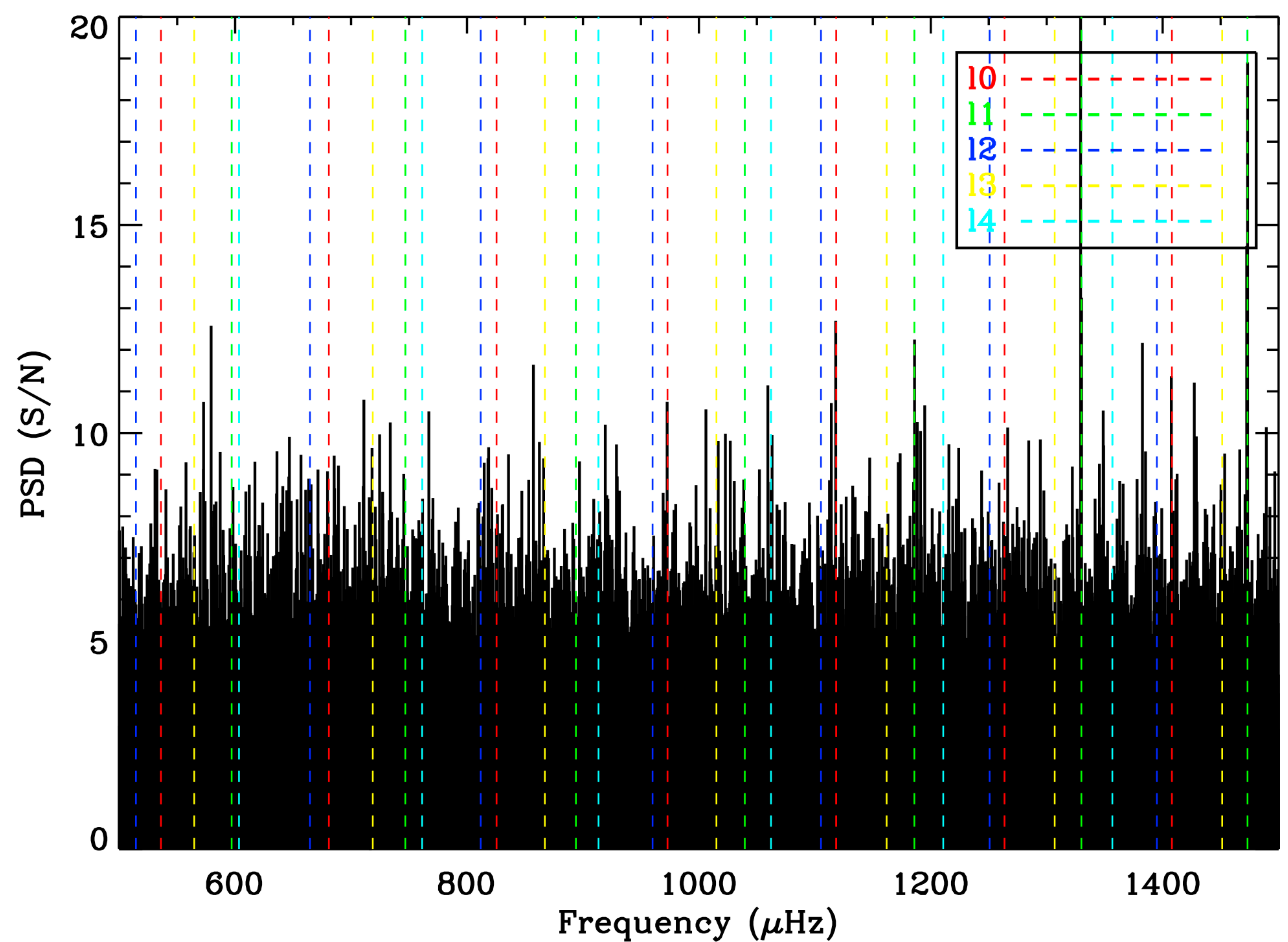}	
\begin{minipage}[b]{14pc}
\caption{\label{Fig0}GOLF PSD in units of signal-to-noise ratio showing the low p-mode frequency region (500 to 1500 $\mu$Hz) computed from continuous 6000 days velocity time series starting on April, 11 1996.}
\end{minipage} 
\end{figure}

GOLF is a resonant scattering spectrophotometer measuring the Doppler wavelength shift - integrated over the solar surface -  of the D1 and D2 Fraunhofer sodium lines at 589.6 and 589.0 nm respectively \cite{Gar98}). The instrument was also designed to measure the solar mean magnetic field of the Sun \cite{Gar99}. Unfortunately, due to occasional malfunctions of the polarization mechanisms switching between both wings of the Na profile and between the two sigma components of the solar light, the instrument was operated in a single-wing configuration in early 1996 \cite{Gar05}.
The low instrumental noise, the fast sampling rate, and the high stability of the platform orbiting around the Lagrangian L1 point allows GOLF to measure: a) the oscillation properties of the Sun at  very low frequencies, allowing the first detection of the asymptotic properties of dipole g modes \cite{Gar07, Gar08} and the dynamics of the radiative interior \cite{Gar04,Mat08}; b) the low-order p modes at around 1 mHz \cite{Ber00, Gar01}; c)  the pseudomodes region above the cut-off frequency \cite{Gar98b}. In Fig.~\ref{Fig0} the low-frequency range between 500 and 1500 $\mu$Hz is shown. 
Some non previously reported p modes seem to rise from the background, e.g. the l=0, n=7 at around 1118 $\mu$Hz.

\begin{figure}[h]
\begin{minipage}{18pc}
\includegraphics[width=18pc]{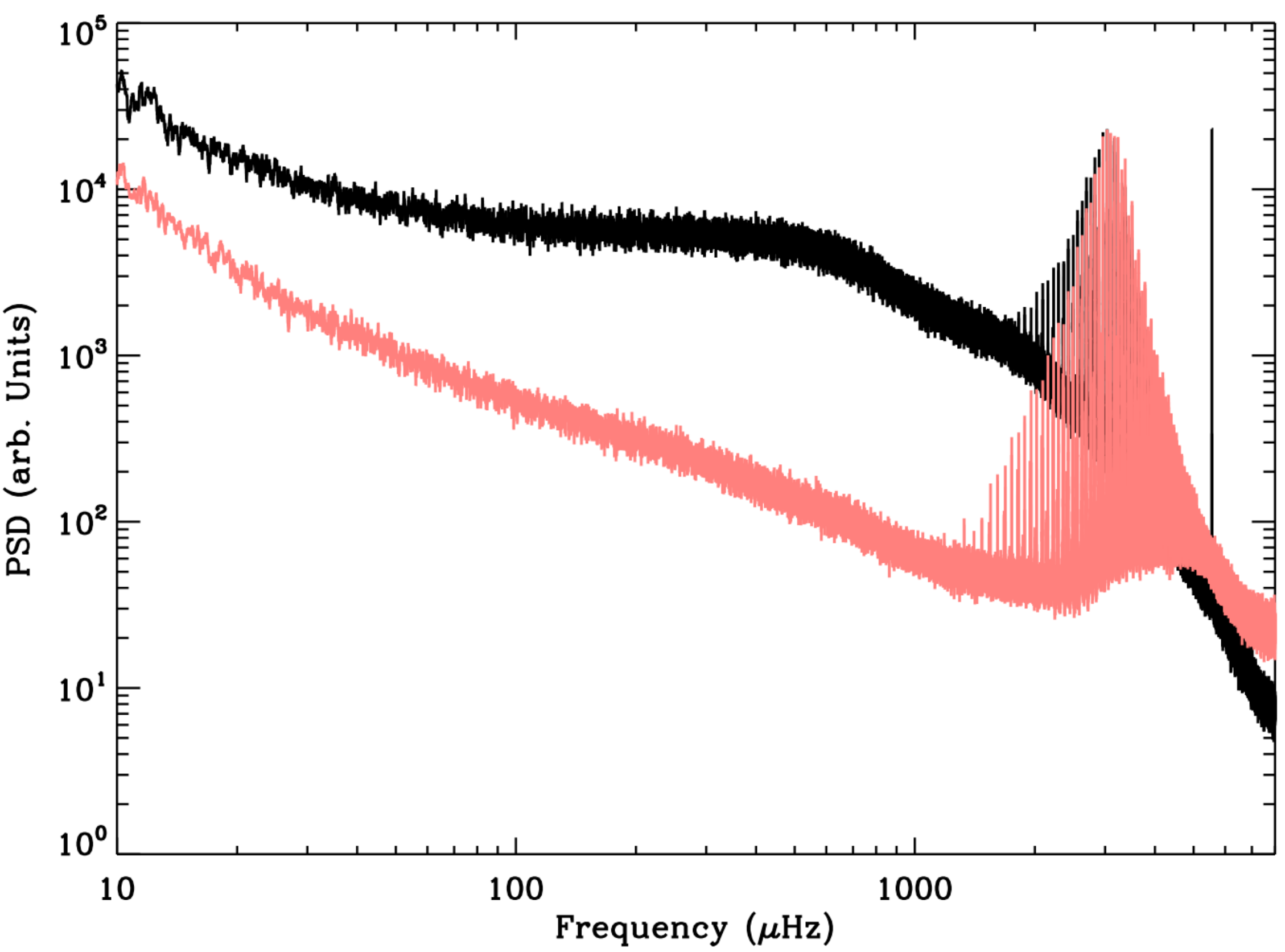}
\caption{\label{Fig1} Power Spectrum Density (PSD) of GOLF (magenta) and VIRGO/SPM green channel (black). Both PSDs have been normalized by the maximum of the p-mode hump and smoothed by a boxcar of 100 bins. The spike at 5555 $\mu$Hz is due to the period of the calibration reference used by the VIRGO/SPM Data Acquisition System.}
\end{minipage}\hspace{1pc}%
\begin{minipage}{18pc}
\includegraphics[width=18pc]{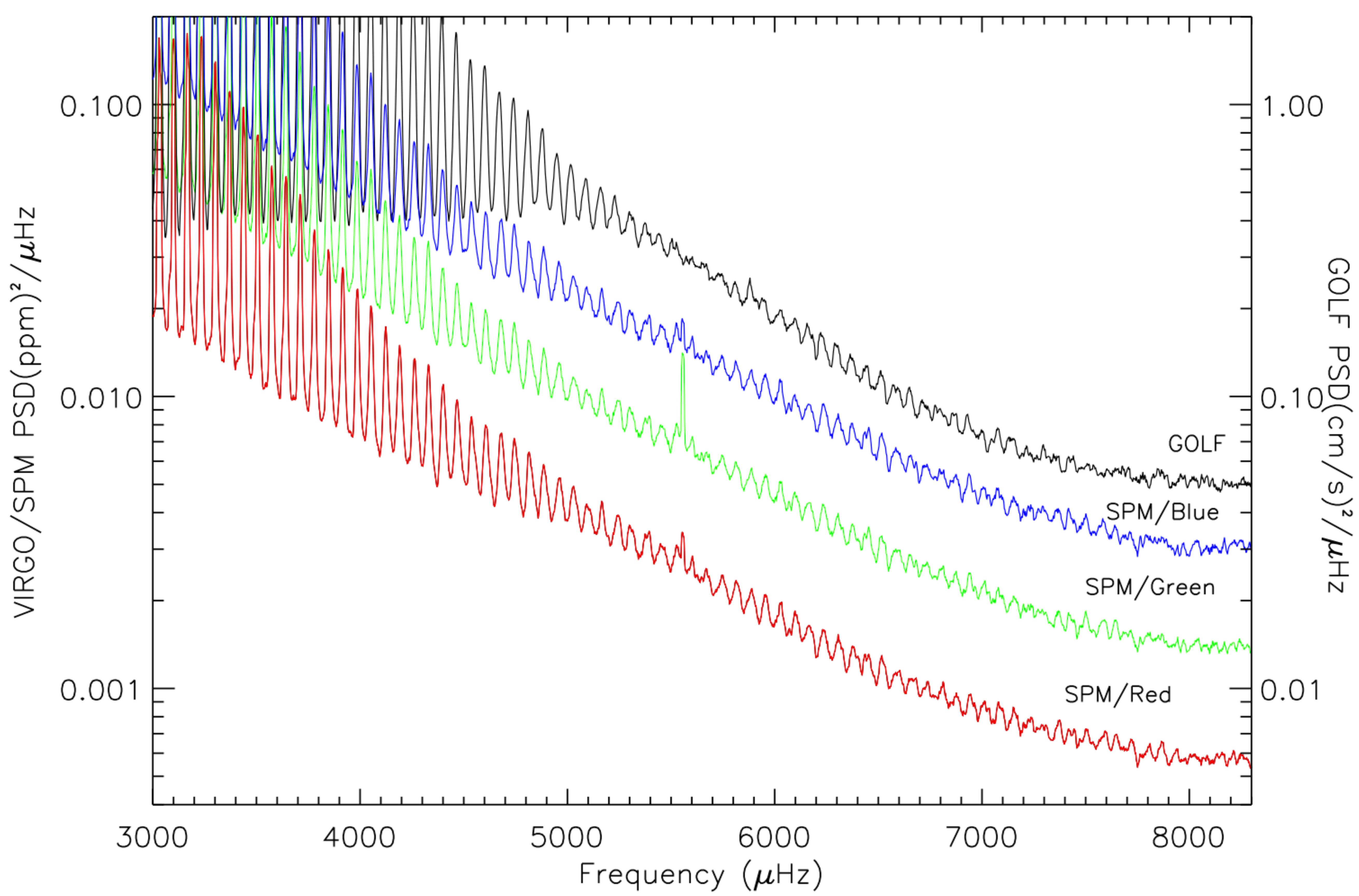}
\caption{\label{Fig2} High-frequency region of the Power Spectrum Density of GOLF and the three VIRGO/SPM. The pseudo-mode region is clearly visible above the cut-off frequency.}
\end{minipage} 
\end{figure}

VIRGO is composed of three type of instruments: two active-cavity radiometers, three Sun spectrophotomenters (SPM) at 402 nm (Blue), 500 nm (Green), and 862 nm (Red), and one Low resolution imager (LOI). Although the solar convective background is higher at lower frequencies compared to velocity measurements (see Fig~\ref{Fig1}), 
VIRGO/SPM data have been very useful to study solar oscillation properties from $\sim 2$ mHz up to the cut-off frequency, including also the pseudo-modes region  \cite{Ajm05} (see Fig.\ref{Fig2}). A comparison of the p-mode properties measured in velocity and intensity can be found at \cite{Tou97}.

\section{The BiSON network}

\begin{figure}[!htb]
\includegraphics[width = 0.70\textwidth]{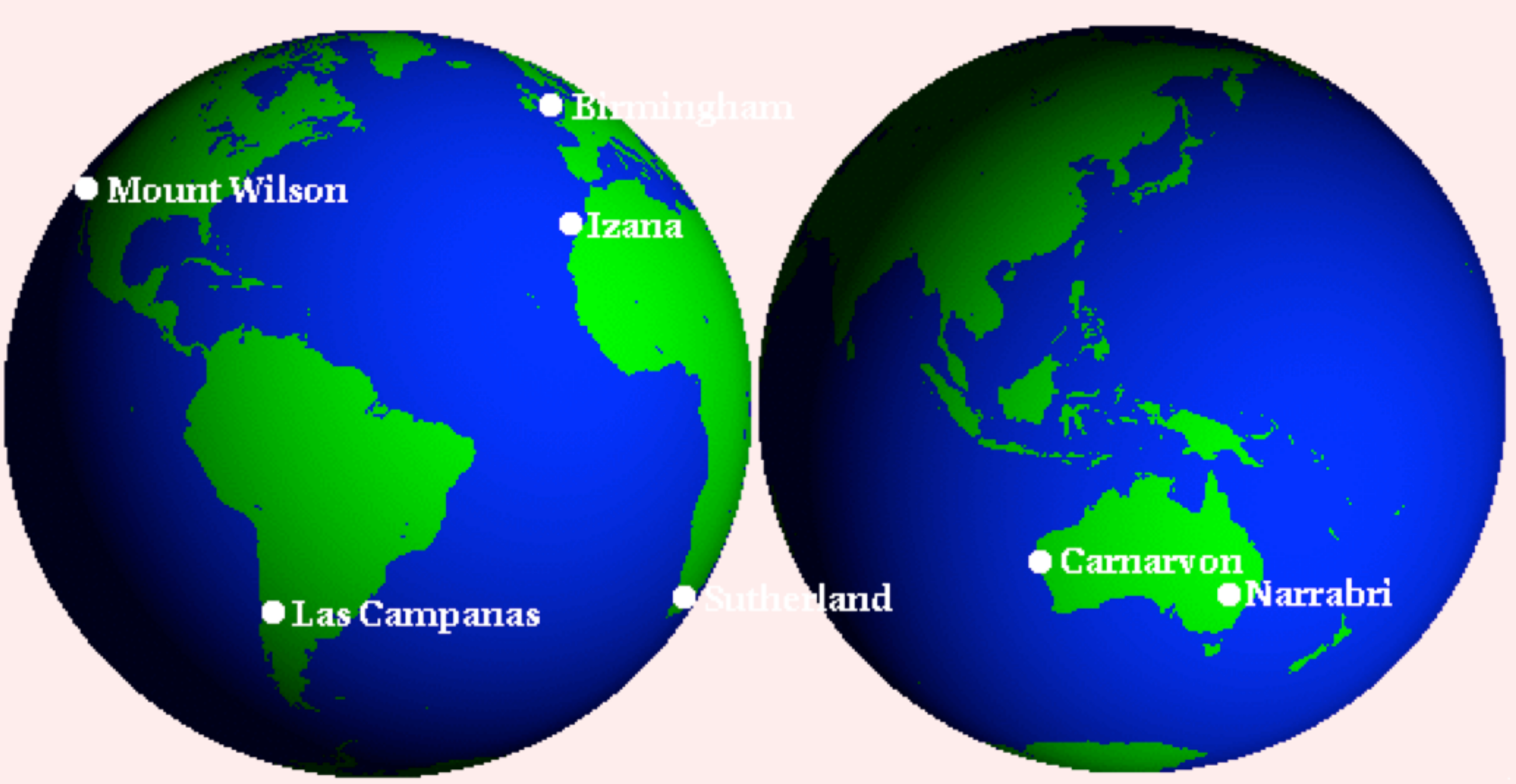}	
\begin{minipage}[b]{10.5pc}
\caption{\label{Fig3}Distribution of the BiSON sites all over the world. }
\end{minipage}
\end{figure}

The Birmingham Solar Oscillation Network (BiSON) is a 6-station network capable of observing the Sun 24h a day (see Fig.~\ref{Fig3}). Although many technical improvements in design have been implemented in the newer stations, the fundamental principles of the Resonant Scattering Spectrometer are unchanged  \cite{Bro76}.

\begin{figure}[!htb]
\includegraphics[width = 0.6\textwidth]{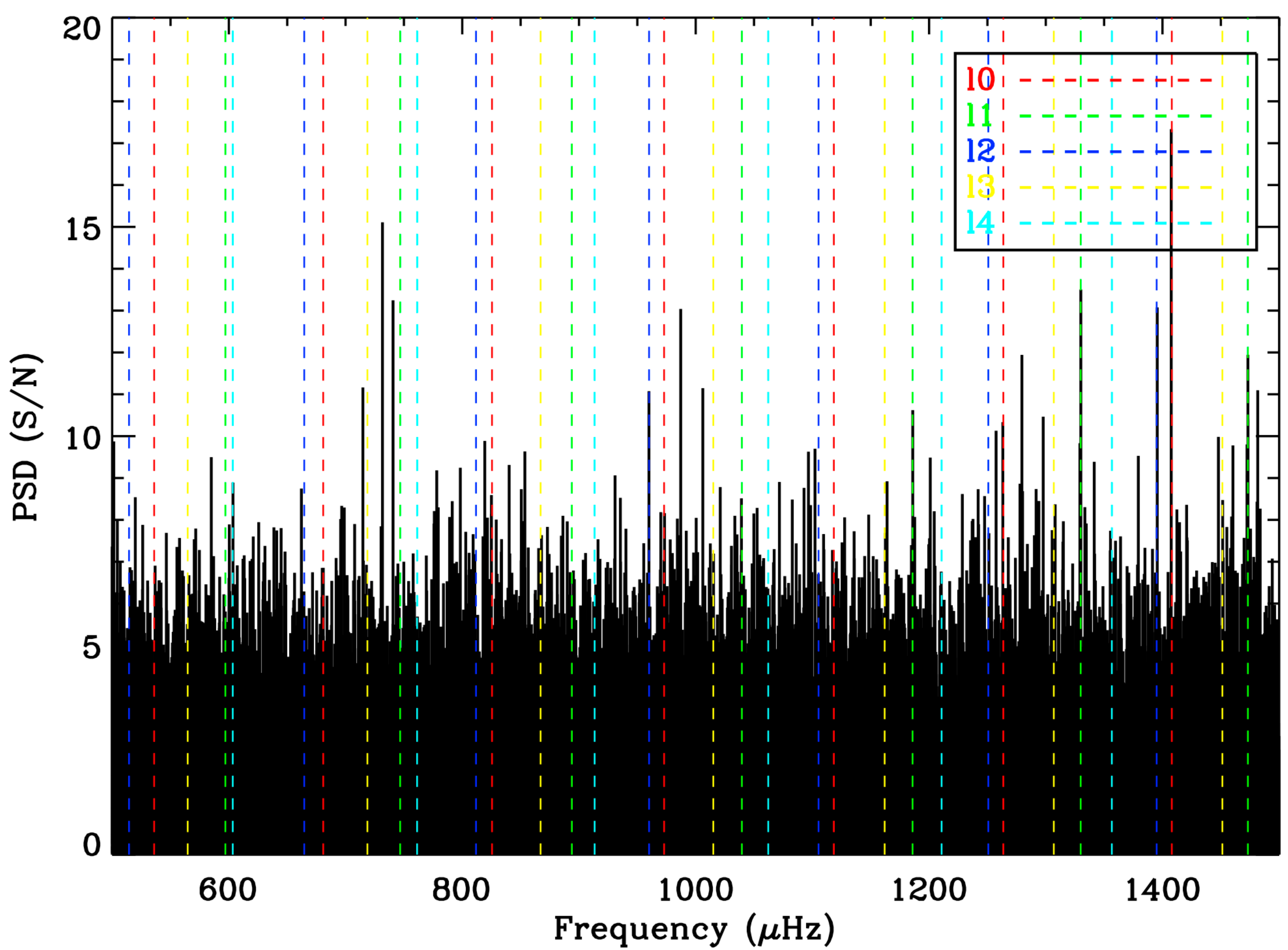}	
\begin{minipage}[b]{14pc}
\caption{\label{Fig5} BiSON PSD expressed in units of signal-to-noise ratio in the low-frequency regime (500 to 1500 $\mu$Hz).  It corresponds to a velocity time series of 2800 days starting January 1, 1995 with an overall duty cycle of 71$\%.$}
\end{minipage} 
\end{figure}

BiSON instruments measure the Doppler-velocity shifts in the K Fraunhofer line at 770 nm. Near-continuous observations have been taken since July 31st, 1978 \cite{Cla79}, with year-by-year duty cycles regularly exceeding 80\% since the network was completely deployed. BiSON is a unique tool for the study of the changes in oscillation frequencies that accompany the solar cycle because it has been collecting data for the past 30 yrs. BiSON observations cover most of cycle 21, the complete cycles 22 and 23, also cycle 24 to present \cite{Bas12}. An example of the BiSON spectrum is shown in Fig.~\ref{Fig5}.

\ack
SoHO is a space mission of international cooperation between ESA and NASA. R.A.G., S.T-C., and D.S. thank the support from CNES. This work was supported in part by the NASA grant NNX09AE59G, by the White Dwarf Research Corporation through the Pale Blue Dot project, and by the grant AYA2010-17803 from the Spanish National Research Plan. NCAR is supported by the National Science Foundation. BiSON is funded by the UK Science Technology and Facilities Council (STFC).


\end{document}